\documentclass{CSML}
\pdfoutput=1

\usepackage{lastpage}
\newcommand{\dOi}{14(2:11)2018}
\lmcsheading{}{1--\pageref{LastPage}}{}{}%
{Apr.~05, 2017}{May~22, 2018}{}


\usepackage{hyperref}
\hypersetup{hidelinks}
\usepackage{amsmath}
\usepackage{amssymb}

\newcommand{\dom}{\operatorname{dom}}

\newcommand{\id}{\textnormal{id}}
\newcommand{\Cantor}{\{0, 1\}^\mathbb{N}}
\newcommand{\Baire}{{\mathbb{N}^\mathbb{N}}}
\newcommand{\uint}{{[0,1]}}

\newcommand{\hide}[1]{}
\newcommand{\mto}{\rightrightarrows}

\newcommand{\mlr}{\textrm{MLR}}

\newcommand{\C}{\textrm{C}}
\newcommand{\lpo}{\textrm{LPO}}
\newcommand{\name}[1]{\textsc{#1}}
\newcommand{\leqW}{\leq_{\textnormal{\textrm{W}}}}

\newcommand{\nleqW}{\nleq_{\textrm{W}}}

\newcommand{\leW}{<_{\textrm{W}}}
\newcommand{\equivW}{\equiv_{\textnormal{\textrm{W}}}}

\newcommand{\lay}{\textnormal{\textrm{LAY}}}
\newcommand{\rd}{\textnormal{\textrm{RD}}}
\newcommand{\kol}{\textrm{Kol}}
\newcommand{\lil}{\textrm{LIL}}

\begin{document}

\title{Weihrauch-completeness for layerwise computability}

\author[A.~Pauly]{Arno Pauly\rsuper{1}}
\address{\lsuper{1}Department of Computer Science\\ Swansea University, United Kingdom
 \\ \& \\ School of Computer Science \\ University of Birmingham, United Kingdom}
\email{Arno.Pauly@cl.cam.ac.uk}
\thanks{During the conception of this work, Pauly was at the University of Cambridge.}

\author[W.~Fouch\'e]{Willem Fouch\'e\rsuper{2}}
\author[G.~Davie]{George Davie\rsuper{2}}
\address{\lsuper{2}Department of Decision Sciences\\ School of Economic Sciences, University of South Africa, SA}
\email{fouchwl@unisa.ac.za}
\email{Davieg@unisa.ac.za}

\begin{abstract}
We introduce the notion of being Weihrauch-complete for layerwise computability and provide several natural examples related to complex oscillations, the law of the iterated logarithm and Birkhoff's theorem. We also consider hitting time operators, which share the Weihrauch degree of the former examples but fail to be layerwise computable.
\end{abstract}

\maketitle

\section{Introduction}
\emph{Layerwise computability} is an effective counterpart to continuous functions that are almost-everywhere defined. This notion was introduced by \name{Hoyrup} and \name{Rojas} \cite{hoyrup4}. A function defined on Martin-L\"of random inputs is called layerwise computable if it becomes computable if each input is equipped with some bound on the layer where it passes a fixed universal Martin-L\"of test. Interesting examples of functions that are layerwise computable but not computable are obtained e.g.~from Birkhoff's theorem or the study of algorithmically random Brownian motion (more below).

Weihrauch reducibility \cite{brattka2,brattka3} is a framework to compare the extent of non-computability of multivalued functions. It has been proposed with a meta-mathematical investigation of the constructive content of existence theorems in mathematics in mind. However, it has also been fruitfully employed to study (effective) function classes such as (effective) Borel measurability \cite{paulybrattka} or piecewise continuity (computability) and (effective) $\Delta^0_2$-measurability \cite{paulydebrecht}.

Our interest in this paper is in problems that are Weihrauch-complete for layerwise computability, i.e.~problems that are layerwise computable, and every layerwise computable problem is Weihrauch reducible to it. These are, in a sense, those problems where \emph{being layerwise computable} cannot be improved to a stronger computability notion. We shall exhibit several natural examples of problems that are Weihrauch-complete for layerwise computability.

The interaction of layerwise computability and Weihrauch reducibility has also been investigated by \name{H\"olzl} and \name{Shafer} \cite{shafer2}, largely in an independent development.

\section{Background}
We give a very brief introduction to the required concepts from randomness theory (in particular, layerwise computability) and Weihrauch reducibility. A standard reference for randomness is \cite{nies}. Layerwise computability was introduced in \cite{hoyrup4}. A survey of Weihrauch reducibility is available as \cite{pauly-handbook}. This reference also provides a more detailed account of the motivation behind Weihrauch reducibility, and a development of the basic operations on Weihrauch degrees.

\subsection{Weihrauch reducibility}
We recall that a represented space $\mathbf{X} = (X, \delta_\mathbf{X})$ is given by a set $X$ and a partial surjection $\delta_\mathbf{X}: \subseteq \Baire \to X$ onto it. A multivalued function between represented spaces $\mathbf{X} = (X, \delta_\mathbf{X})$ and $\mathbf{Y} = (Y, \delta_\mathbf{Y})$ is a left-total relation between $X$ and $Y$, relating inputs from $\mathbf{X}$ with correct solutions from $\mathbf{Y}$. We write $f : \mathbf{X} \mto \mathbf{Y}$ for this, and use $f : \subseteq \mathbf{X} \mto \mathbf{Y}$ to express that $f$ is a partial multivalued function. A partial function $F : \subseteq \Baire \to \Baire$ is a \emph{realizer} of a multivalued function $f : \mathbf{X} \mto \mathbf{Y}$ (in symbols $F \vdash f$), if $\delta_\mathbf{Y}F(p) \in f(\delta_\mathbf{X}(p))$ for all $p \in \dom(\delta_\mathbf{X})$. A multivalued function between represented spaces is called computable respectively continuous iff it has some computable respectively continuous realizer. Let $\langle \ , \ \rangle : \Baire \times \Baire \to \Baire$ be a standard pairing function.

Given some represented spaces $\mathbf{X}$, $\mathbf{Y}$ we obtain the represented space $\mathcal{C}(\mathbf{X},\mathbf{Y})$ of continuous functions from $\mathbf{X}$ to $\mathbf{Y}$ by fixing a universal oracle Type-2 machine $\Phi$, and letting $q$ be a name for $f : \mathbf{X} \to \mathbf{Y}$ iff $p \mapsto \Phi^q(p)$ is a realizer of $f$. This makes all the usual operations (in particular function application) computable. We introduce the Sierpi\'nski-space $\mathbb{S} := (\{\top,\bot\}, \delta_\mathbb{S})$ where $\delta_\mathbb{S}(0^\mathbb{N}) = \bot$ and $\delta_\mathbb{S}(p) = \top$ if $p \neq 0^\mathbb{N}$. Then we can define the hyperspace $\mathcal{O}(\mathbf{X})$ of open sets by identifying a subset $U \subseteq \mathbf{X}$ with its characteristic function $\chi_U \in \mathcal{C}(\mathbf{X},\mathbb{S})$. For the hyperspace $\mathcal{A}(\mathbf{X})$ of closed sets, we identify a subset $U \subseteq \mathbf{X}$ with the characteristic function of its complement. For details, see \cite{pauly-synthetic}.

Two of these hyperspaces are particularly relevant for us: Regarding $\mathcal{O}(\Cantor)$, we can envision a set $U \in \mathcal{O}(\Cantor)$ to be given by a (finite or infinite) list of finite prefixes $(w_i)_{i \in I}$ such that $U = \bigcup_{i \in I} w_i\Cantor$. Regarding $\mathcal{A}(\mathbb{N})$, we can consider $A \in \mathcal{A}(\mathbb{N})$ to be given by some $p \in \Baire$ such that $n \notin A \Leftrightarrow \exists i \ p(i) = n+1$.

Now we shall introduce Weihrauch reducibility as a preorder on multivalued functions between represented spaces. Intuitively, $f$ being Weihrauch reducible to $g$ means that there is an otherwise computable procedure to solve $f$ by invoking an oracle for $g$ exactly once. We thus obtain a very fine-grained picture of the relative strength of the multivalued functions. Consequently, a Weihrauch equivalence is a very strong result compared to other approaches that allow more generous access to the principle being reduced to.

\begin{defi}[Weihrauch reducibility]
\label{def:weihrauch}
Let $f,g$ be multi-valued functions on represented spaces.
Then $f$ is said to be {\em Weihrauch reducible} to $g$, in symbols $f\leqW g$, if there are computable
functions $K,H:\subseteq\Baire\to\Baire$ such that $\left(p \mapsto K\langle p, GH(p) \rangle \right )\vdash f$ for all $G \vdash g$.
\end{defi}
The relation $\leqW$ is reflexive and transitive. We use $\equivW$ to denote equivalence regarding $\leqW$,
and by $\leW$ we denote strict reducibility.

Products of represented spaces can be defined in the natural way based on $\langle \ , \ \rangle$, and we obtain products of (multivalued) functions between them accordingly. The Weihrauch degree of $f \times g$ depends only on the Weihrauch degrees of $f$ and $g$, i.e.~$\times$ lifts to an operation on Weihrauch degrees as observed in \cite{paulyreducibilitylattice,brattka2}. While $\times$ is the most important operation on Weihrauch degrees in this paper, in Section \ref{sec:thedegree} we will mention two further operations that also correspond to a logical \emph{and}: Access to $f \sqcup g$ means that we can chose to either make a query to $f$ or a query to $g$; while access to $f \star g$ means we can first make a query to $g$, and then (knowing the result) make a query to $f$. The Weihrauch degrees are a lattice, and $\sqcup$ is the join of that lattice. The \emph{parallelization} of $f$, denoted by $\widehat{f}$, gives access to countably many instances of $f$ in parallel. We refer the interested reader to \cite{paulybrattka4} for a detailed investigation of the algebraic structure of the Weihrauch degrees.

A Weihrauch degree that is very relevant for our investigation is closed choice on the natural numbers.
\begin{defi}
Let $\C_{\mathbb{N}} : \subseteq \mathcal{A}(\mathbb{N}) \mto \mathbb{N}$ be defined via $n \in \C_\mathbb{N}(A)$ iff $n \in A$.
\end{defi}

This degree has received significant attention, e.g.~in \cite{brattka3,paulybrattka,paulymaster,mylatz,mylatzb,hoelzl,paulyneumann}. In particular, as shown in \cite{paulydebrecht}, a function between computable Polish spaces is Weihrauch reducible to $\C_\mathbb{N}$ iff it is piecewise computable iff it is effectively $\Delta^0_2$-measurable. For our purposes, the following representatives of the degree are also relevant.

\begin{lem}
\label{lemma:cn}
The following are Weihrauch equivalent:
\begin{enumerate}
\item $\C_\mathbb{N}$
\item $\textrm{UC}_\mathbb{N}$, defined via $\textrm{UC}_\mathbb{N} = \left ( \C_\mathbb{N} \right)|_{\{A \in \mathcal{A}(\mathbb{N}) \mid |A| = 1\}}$
\item $\min : \subseteq \mathcal{A}(\mathbb{N}) \to \mathbb{N}$
\item $\max : \subseteq \mathcal{O}(\mathbb{N}) \to \mathbb{N}$
\item $\operatorname{Bound} : \subseteq \mathcal{O}(\mathbb{N}) \mto \mathbb{N}$, where $n \in \operatorname{Bound}(U)$ iff $\forall m \in U \ n \geq m$.
\end{enumerate}
\begin{proof}
\begin{description}
\item[$1. \equivW 2.$] This is from \cite{paulybrattka}.
\item[$1. \leqW 3.$] Trivial.
\item[$3. \leqW 4.$] Given $A \in \mathcal{A}(\mathbb{N})$, we can compute $U_{\leq A} := \{n \in \mathbb{N} \mid \forall m \in A \ n \leq m\} \in \mathcal{O}(\mathbb{N})$. Now $(\max U_{\leq A}) = \min A$.
\item[$4. \leqW 1.$] If $U \in \dom(\max)$, then $U \neq \emptyset$. Thus, we can assume $U$ to be given as $U = \{p_U(n) \mid n \in \mathbb{N}\}$ for some $p \in \Baire$. Now $A := \{n \in \mathbb{N} \mid \forall m \in \mathbb{N} \ p(m) \leq p(n)\}$ can be computed as a closed set. Applying $\C_\mathbb{N}$ to $A$ to obtain some element $k$, and then computing $p(k)$ yields $\max U$.
\item[$2. \leqW 5.$] As before, we use $U_{\leq A}$, this time on some $A = \{n\}$. Any bound $b$ for $U_{\leq A}$ also is a bound for $n$. We then simply wait until we have learned $k \notin \{n\}$ for all but one $k \leq n$ -- the remaining candidate is the answer to $\textrm{UC}_\mathbb{N}$.
\item[$5. \leqW 4.$] Trivial.
\qedhere
\end{description}
\end{proof}
\end{lem}

We also require the following family of Weihrauch degrees:

\begin{defi}
Given some set $A \subseteq \Baire$, let $d_A : A \to \{1\}$ be the unique map of that type.
\end{defi}

It was shown in \cite{paulykojiro} that $d_{(\cdot)}$ is a lattice embedding of the dual of the Medvedev degrees into the Weihrauch degrees. In particular, we have that $d_A \leqW d_B$ iff there is a computable function $F : A \to B$.

In Section \ref{sec:hitting}, we also mention the degree of $\lpo : \Baire \to \{0,1\}$ where $\lpo(0^\mathbb{N}) = 1$ and $\lpo(p) = 0$ for $p \neq 0^\mathbb{N}$ which was introduced in \cite{weihrauchc}, and the Kleene star operation $^*$ from \cite{paulyreducibilitylattice,paulyincomputabilitynashequilibria} defined by $f^{0} := \id_\Baire$, $f^{n+1} := f^{n} \times f$ and $f^*(n,x) := f^n(x)$.

\subsection{Randomness}
Let $\lambda$ denote the standard Lebesgue measure on $\Cantor$. A \emph{Martin-L\"of test} in $\Cantor$ is a computable sequence $(U_i)_{i \in \mathbb{N}}$ of open sets such that $\lambda(U_i) \leq 2^{-i}$. A Martin-L\"of test $(U_i)_{i \in \mathbb{N}}$ is called \emph{universal}, if for any Martin-L\"of test $(V_i)_{i \in \mathbb{N}}$ we find that $\left (\bigcap_{i \in \mathbb{N}} V_i \right ) \subseteq  \left (\bigcap_{i \in \mathbb{N}} U_i \right )$. Universal Martin-L\"of tests exist, and we call $\mlr := \Cantor \setminus \left (\bigcap_{i \in \mathbb{N}} U_i \right )$ for some universal Martin-L\"of test the set of \emph{Martin-L\"of random sequences}. The set $\mlr$ is independent of the choice of the universal test.

The informal idea behind Martin-L\"of randomness is that a Martin-L\"of test $(U_i)_{i \in \mathbb{N}}$ describes a very specific computable property $\left (\bigcap_{i \in \mathbb{N}} U_i \right )$, and that a random sequence should not have any very specific computable properties. Note that for any Martin-L\"of test $(U_i)_{i \in \mathbb{N}}$, also $\left (\bigcap_{i \leq n} U_i \right )_{n \in \mathbb{N}}$ is a Martin-L\"of test describing the same property. Thus, nothing substantial would change if we would require $U_{i+1} \subseteq U_i$ to hold in any test, i.e.~would require the tests to be \emph{nested}\footnote{Which in fact was part of the original definition by \name{Martin-L\"of} \cite{martinloef}. Considering also non-nested tests though adds potential expressivity to the concept of layerwise computability, below.}.

Following \cite{miyabe}, a Martin-L\"of test $(U_i)_{i \in \mathbb{N}}$ is called \emph{optimal}, if for  any Martin-L\"of test $(V_i)_{i \in \mathbb{N}}$ we find that  there is some $n \in \mathbb{N}$ such that $\forall i \in \mathbb{N} \ V_{i + n} \subseteq U_{i}$. Note that any optimal Martin-L\"of test is necessarily universal. The existence of optimal Martin-L\"of tests was established in \cite{miyabe}.

A function $f : \mlr \to \mathbf{X}$ is called \emph{layerwise computable} (w.r.t.~the universal test $(U_i)_{i \in \mathbb{N}}$), if there is a computable function $g : \subseteq \mlr \times \mathbb{N} \to \mathbf{X}$ such that $p \notin U_k \Rightarrow g(p,k) = f(p)$. As shown in \cite{shafer2}, the notion of layerwise computability does depend on the choice of universal test. If a function is layerwise computable for some universal test, then it is layerwise computable for any optimal test. We extend the notion of layerwise computability to multivalued functions $f : \mlr \mto \mathbf{X}$, by considering computable multivalued $g : \subseteq \mlr \times \mathbb{N} \mto \mathbf{X}$ such that $p \notin U_k \Rightarrow \emptyset \neq g(p,k) \subseteq f(p)$.

An alternate (but equivalent) approach to randomness is expressed in terms of Kolmogorov complexity. We fix a prefix-free universal Turing machine, and then let $K(w)$ be the length of the shortest programme computing the string $w \in \{0,1\}^*$. For $p \in \Cantor$ and $n \in \mathbb{N}$, let $p_{\leq n}$ be the prefix of $p$ of length $n$. Then for $c \in \mathbb{N}$ we set $\textrm{K}^d = \{p \in \Cantor \mid \forall n \in \mathbb{N} \ K(p_{\leq n}) \geq n - d\}$, and find that $\mlr = \bigcup_{d \in \mathbb{N}} \textrm{K}^d$. Based on counting the number of prefix-free programs of a certain length, we find that $\lambda((\textrm{K}^d)^C) \leq 2^{-d}$; moreover, each set $(\textrm{K}^d)^C$ is computably open. It is know that $\left ( (\textrm{K}^d)^C \right)_{d \in \mathbb{N}}$ is a universal Martin-L\"of test. For more details, see \cite{nies} for example.

\section{The Weihrauch degree}
\label{sec:thedegree}

\begin{defi}
Fix some universal Martin-L\"of test $\mathcal{U} = (U_n)_{n \in \mathbb{N}}$. Let $\lay_\mathcal{U} : \mlr \mto \mathbb{N}$ be defined via $n \in \lay_\mathcal{U}(p)$ iff $p \notin U_n$. Let $\rd_\mathcal{U} : \mlr \to \mathbb{N}$ be defined via $\rd_\mathcal{U}(p) = \min \{n \in \mathbb{N} \mid p \notin U_n\}$.
\end{defi}

\begin{obs}
\label{obs:layer}
If $f : \mlr \mto \mathbf{X}$ is layerwise computable (w.r.t.~$\mathcal{U}$), then $f \leqW \lay_\mathcal{U}$.
\end{obs}

\begin{thm}
\label{theo:main}
$\lay_\mathcal{U} \equivW \rd_\mathcal{U} \equivW \C_\mathbb{N} \times d_{\mlr}$
\begin{proof}
\begin{description}
\item[$\lay_\mathcal{U} \leqW \rd_\mathcal{U}$] Trivial.
\item[$\rd_\mathcal{U} \leqW \C_\mathbb{N} \times d_{\mlr}$] As $\dom(\rd_\mathcal{U}) = \mlr$, we have a random sequence available as input for $d_\mlr$, and the presence of this degree does not matter further. To see that $\C_\mathbb{N}$ suffices to obtain the answer, note that given $p$ we can compute $\{n \mid p \notin U_n\} \in \mathcal{A}(\mathbb{N})$. By Lemma \ref{lemma:cn}, $\C_\mathbb{N}$ lets us compute the minimum of a closed set.
\item[$\C_\mathbb{N} \times d_{\mlr} \leqW \lay_\mathcal{U}$] By Lemma \ref{lemma:cn}, we may show $\operatorname{Bound} \times d_\mlr$ instead. This works as follows:

    The input is an enumeration of some finite set $I \subset \mathbb{N}$ (which we may safely assume to be an interval) and a random sequence $p$. Let $w$ be the current prefix of the output (i.e.~the input to $\lay_\mathcal{U}$). If we learn that $n \in I$, we consider $w0^\mathbb{N}$. As this is not random and $\mathcal{U}$ is universal, we know that $w0^\mathbb{N} \in U_n$. As $U_n$ is open, there is some -- effectively findable -- $k \in \mathbb{N}$ such that $w0^k\Cantor \subseteq U_n$. We proceed to amend the current output to $w0^k$, and then start outputting $p$ (until we potentially learn $n + 1 \in I$).

    As $I$ is finite, the output $q$ will have some tail identical to $p$, and thus is Martin L\"of random. By construction, whenever $n \in I$, then $q \in U_n$, thus if $b \in \lay_\mathcal{U}(q)$ then $b \in \operatorname{Bound}(p)$.
\qedhere
\end{description}
\end{proof}
\end{thm}

There are a number of important consequences of this result. First, as the right hand side does not depend on the choice of the universal Martin L\"of test, we see that the Weihrauch degree of $\lay_\mathcal{U}$ and $\rd_\mathcal{U}$ is independent of the test, too. Thus, in the following we suppress the subscript $\mathcal{U}$. Further consequences are:

\begin{cor}
\label{corr:star}
$\lay \times \lay \equivW \lay$ and $\lay \star \lay \equivW \lay$.
\begin{proof}
The former statement follows from the latter. For any $A \subseteq \Baire$ and Weihrauch degrees $f, g$, we find that $(d_A \times f) \star (d_A \times g) \equivW d_A \times (f \star g)$. This is because $d_A$ produces no output useful for producing the input of $f$, and the second instance of $d_A$ can be fed the same input as we use for the first. In particular, we have that $(d_\mlr \times \C_\mathbb{N}) \star (d_\mlr \times \C_\mathbb{N}) \equivW d_\mlr \times (\C_\mathbb{N} \star \C_\mathbb{N})$. The independent choice theorem from \cite{paulybrattka} implies that $\C_\mathbb{N} \star \C_\mathbb{N} \equivW \C_\mathbb{N}$. The latter claim now follows from Theorem \ref{theo:main}.
\end{proof}
\end{cor}

\begin{cor}
$\lay \leW \C_\mathbb{N}$.
\begin{proof}
As $d_\mlr$ is computable, we find $d_\mlr \times \C_\mathbb{N} \leqW \C_\mathbb{N}$. That $\C_\mathbb{N} \nleqW d_\mlr \times \C_\mathbb{N}$ follows from the fact that $\C_\mathbb{N}$ has computable inputs, whereas $d_\mlr \times \C_\mathbb{N}$ does not.
\end{proof}
\end{cor}

\begin{cor}
$\lay \star \C_\mathbb{N} \equivW \C_\mathbb{N} \star \lay \equivW \lay$.
\begin{proof}
Same reasoning as for Corollary \ref{corr:star}.
\end{proof}
\end{cor}

Let $\lim : \subseteq (\Baire)^\mathbb{N} \to \Baire$ map a converging sequence to its limit.

\begin{cor}
$\lay \leW \widehat{\lay} \equivW \lim \times d_\mlr$.
\begin{proof}
That $\widehat{\lay} \equivW \lim \times d_\mlr$ follows from $\widehat{\mathbf{a} \times \mathbf{b}} \equivW \widehat{\mathbf{a}} \times \widehat{\mathbf{b}}$ as shown in \cite{paulybrattka4} together with $\widehat{d_\mlr} \equivW d_\mlr$ and $\widehat{\C_\mathbb{N}} \equivW \lim$ as shown in \cite{brattka2}. That $\lim \times d_\mlr \nleqW \lay$ follows from $\lay$ produces only computable outputs whereas $\lim \times d_\mlr$ can produce the Halting problem from arbitrary random degrees, and there are Martin-L\"of random sequences that do not compute the Halting problem.
\end{proof}
\end{cor}

\begin{cor}
$\lay \leW \lay^* \equivW \id_\Baire \sqcup \lay \leW \C_\mathbb{N}$.
\begin{proof}
By iterating Corollary \ref{corr:star} we see that $\left (\lay \right )^{n} \equivW \lay$ for $n > 0$. As the proof is completely uniform, this implies $\lay^* \equivW \id_\Baire \sqcup \lay$. As this degree has a computable point in its domain, we conclude $\lay^* \nleqW \lay$.

Let $\max_c$ be the restriction of $\max : \subseteq \mathcal{O}(\mathbb{N}) \to \mathbb{N}$. The proof of Lemma \ref{lemma:cn} shows that $\max_c \equivW \C_\mathbb{N}$. If $\C_\mathbb{N} \leqW \lay^*$ would hold, then we would also have $\max_c \leqW (\id_\Baire \sqcup \lay)$. However, as $\max_c$ has only computable inputs, we can never produce a valid input for $\lay$ in that putative reduction, and hence see that $\max_c \leqW \id_\Baire$ would follow, i.e.~that $\max_c$ were computable. This is false, hence $\C_\mathbb{N} \nleqW \lay^*$ holds.
\end{proof}
\end{cor}

\begin{cor}
\label{corr:domain}
If $f \leqW \C_\mathbb{N}$ for $f : \subseteq \mlr \mto \mathbf{Y}$, then $f \leqW \lay$.
\end{cor}

\begin{cor}
\label{corr:jaynerogers}
The following are equivalent for $f : \subseteq \mlr \to \mathbf{Y}$ for a computable metric space $\mathbf{Y}$:
\begin{enumerate}
\item $f$ is effectively $\Delta^0_2$-measurable.
\item $f$ is $\Pi^0_1$-piecewise computable.
\item $f \leqW \lay$.
\end{enumerate}
\begin{proof}
This is obtained by combining the computable Jayne-Rogers theorem from \cite{paulydebrecht} with Corollary \ref{corr:domain}.
\end{proof}
\end{cor}

Most results in this section were independently obtained by \name{H\"olzl} and \name{Shafer} in \cite{shafer2}, Corollaries \ref{corr:domain} and \ref{corr:jaynerogers} are inspired by their corresponding results though. The proofs  in \cite{shafer2} differ significantly from ours, in particular, they give direct proofs of the claims listed as corollaries here.

In a very similar fashion to Theorem \ref{theo:main}, we can also characterize the degree of Kolmogorov randomness. While this technically is just a special case of Theorem \ref{theo:main}, we provide a direct proof in the hope to illuminate the underlying phenomena. Let $\kol : \mlr \to \mathbb{N}$ be defined via $\kol(p) := \min \{c \in \mathbb{N} \mid \forall n \in \mathbb{N} \ K(p_{\leq n}) \geq n - c\}$. Then:

\begin{prop}
$\kol \equivW \C_\mathbb{N} \times d_\mlr$
\begin{proof}
Note that $\{c \in \mathbb{N} \mid \forall n \in \mathbb{N} \ K(p_{\leq n}) \geq n - c\}$ can be computed as a closed set from $p$ -- if some $c$ is not in that set, we can find some $n$ and some short program (of length less than $n - c$) producing the prefix $p_{\leq n}$. The reduction $\kol \leqW \C_\mathbb{N} \times d_\mlr$ then follows from Lemma \ref{lemma:cn}.

For the other direction, we show $\operatorname{Bound} \times d_\mlr \leqW \kol$ and again invoke Lemma \ref{lemma:cn}. Given some $w \in \{0,1\}^n$ and $k \in \mathbb{N}$, there will be some programme for our fixed universal machine printing $w0^k$ of size $O(n \log k)$. Based on the constant involved, $n$ and $c$ we can choose $k$ sufficiently large such that $K(w0^k) + c < n + k$.

Now our reduction works as follows: Copy the random sequence serving as the input to $d_\mlr$ over to the input for $\kol$. Whenever we learn that some $c$ is in the input to $\operatorname{Bound}$, we pick a $k$ based on the current prefix of the input to $\kol$ and $c$ and write the corresponding number of zeros. Then we continue to copy the random sequence. Eventually the input to $\operatorname{Bound}$ stabilizes, so our input for $\kol$ will actually be random. Moreover, by constructing, the output of $\kol$ will exceed all numbers in the input to $\operatorname{Bound}$.
\end{proof}
\end{prop}

\section{Examples of Weihrauch-complete layerwise computable operations}
\subsection{Complex oscillations}
Let $\mathcal{C}_0(\uint, \mathbb{R})$ denote the space of continuous functions $f : \uint \to \mathbb{R}$ where $f(0) = 0$. The complex oscillations $\textrm{CO}$ (introduced in \cite{asarin}) are the Martin-L\"of random elements (in the sense of \cite{hoyrup5}) of $\mathcal{C}_0(\uint, \mathbb{R})$ equipped with the Wiener measure. They are of great interest as generic representatives of Brownian motion \cite{fouche2}. We shall consider a specific bijection $\Phi : \mlr \to \textrm{OC}$ studied in \cite{fouche2}.

The definition of $\Phi$ is as follows:

\begin{equation}
\Phi(\alpha)(t)=g(\alpha _{0})\Delta _{0}(t)+g(\alpha_{1})\Delta
_{1}(t)+\sum_{j\geq 1}\sum_{n<2^{j}}g(\alpha _{jn})\Delta _{jn}(t)\text{.}
\tag{4}
\end{equation}%
The $\Delta _{0}(t),\Delta _{1}(t),\Delta _{jn}(t)$ are the sawtooth
functions obtained by integrating from $0$ to $t$ the elements of the Haar
system of functions,
\begin{align*}
  e_{0}&=1,\\
  e_{1}&=\chi ([0,\frac{1}{2}))-\chi ([\frac{1}{2},1)),\\
  e_{jn}&=\{\chi ([n2^{-j},n2^{-j}+2^{-(j+1)}))-\chi ([n2^{-j}+2^{-(j+1)},(n+1)2^{-j}))\}2^{j/2},
\end{align*}
$0\leq n<2^{j}$ and $j\geq 1$. 

The function $g$ is implicitly defined to satisfy $\alpha =\int_{-\infty }^{g(\alpha)}\frac{e^{-t^{2}/2}}{%
\sqrt{2\pi }}dt$ for $\alpha \in (0,1)$. The numbers $\alpha _{0}$,$\alpha_{1}$,$\alpha _{jn}$ are obtained by partitioning the sequence $\alpha$ appropriately into disjoint subsequences, and interpreting these as binary expansions of real numbers from $\uint$. The details of the partitioning do not matter for our purposes, only that it is done in a computable fashion. See \cite{fouche2} for details. We now set $\eta _{0}=g(\alpha_{0}),\eta _{1}=g(\alpha _{1})$ and $\eta
_{jn}=g(\alpha _{jn}).$ By construction, these are independent $\mathcal{N}(0,1)$ random variables
w.r.t. the Lebesgue measure.




We shall require the basic:

\begin{fact}
There is a computable function $\eta : \mlr \to \mathbb{R}$ inducing the normal distribution $\mathcal{N}(0,1)$ on $\mathbb{R}$, in the sense that for any Borel set $A \subseteq \mathbb{R}$ the measure assigned to $A$ according to $\mathcal{N}(0,1)$ is equal to $\lambda(\eta^{-1}(A))$. \footnote{As shown in \cite{schroder2}, this is just saying that $\mathcal{N}(0,1)$ is a computable probability measure.}
\end{fact}

\begin{obs}
$\max : \mathcal{C}(\uint, \mathbb{R}) \to \mathbb{R}$ and $\operatorname{GreaterNat} : \mathbb{R} \mto \mathbb{N}$ where $n \in \operatorname{GreaterNat}(x)$ if $x \leq n$ are computable.
\begin{proof}
For the former, see e.g.~\cite[Corollary 10.9]{pauly-synthetic}. The latter is trivial.
\end{proof}
\end{obs}

\begin{lem} \cite{fouche2}
The  function $\Phi : \mlr \to \textrm{CO}$ can be recursively defined from the values $\Phi(\alpha)$ takes on the dyadic rationals, and then extending it continuously to the interval. To wit:
\begin{enumerate}
\item $\Phi(\alpha)(1) := \eta(\alpha_0)$
\item $\Phi(\alpha)(\frac{1}{2}) := \frac{1}{2} \left ( \eta(\alpha_0) + \eta(\alpha_1) \right )$
\item $\Phi(\alpha)(\frac{2n+1}{2^{j+1}}) := \frac{1}{2} \left ( 2^{-j/2} \eta(\alpha_{jn}) + \Phi(\alpha)(\frac{n+1}{2^j}) + \Phi(\alpha)(\frac{n}{2^j}) \right )$
\end{enumerate}
\end{lem}

\begin{lem}
\label{lemma:coaux}
Given $k \in \mathbb{N}$ and $v \in \{0,1\}^*$ we can compute some $w \in \{0,1\}^*$ such that for all $\alpha \in \mlr$ we find that $k < \sup_{t \in \uint} \Phi(vw\alpha)(t)$.
\begin{proof}
Pick some $j, n \in \mathbb{N}$ such that $\alpha_{jn}$ in $\alpha = \langle \alpha_0, \alpha_1, \ldots, \alpha_{jn}, \ldots\rangle$ does not depend on the prefix of length $|v|$ at all. We can then choose a prefix of $\alpha_{jn}$ (and prefixes of the $\alpha_{j'n'})$ to enforce that $\eta(\alpha_{jn})$ is large enough to ensure that $\Phi(\beta)(\frac{2n+1}{2^{j+1}}) > k$ for all $\beta$ sharing these prefixes. From these prefixes, we obtain $w$.
\end{proof}
\end{lem}

\begin{thm}
$\Phi \equivW \lay$.
\begin{proof}
It was shown in \cite{davie2} that $\Phi$ is layerwise computable. We sketch the argument: From the definition of $\Phi(\alpha)$ we learn how to compute the values taken by $\Phi(\alpha)$ on a dense subset. To obtain $\Phi(\alpha)$ as an element in $\mathcal{C}(\uint, \mathbb{R})$, we also need a modulus of continuity. In \cite{fouche3}, it is shown that the following holds for sufficiently small $h$:
$$\sup_{t \in \uint} |\Phi(\alpha)(t + h) - \Phi(\alpha)(t)| < \sqrt{3h\log(h^{-1})}$$
By inspecting the proof we see that knowing a bound for the layer of $\alpha$ suffices to determine what \emph{small enough} means for $h$ -- and then we have a modulus of continuity.

That $\Phi \leqW \lay$ then follows immediately by Observation \ref{obs:layer}, so it only remains for us to show $\lay \leqW \Phi$. By Theorem \ref{theo:main} and Lemma \ref{lemma:cn}, we can show $d_\mlr \times \operatorname{Bound} \leqW \Phi$ instead. For that, we describe how we compute an input to $\Phi$ from inputs to $d_\mlr$ and $\operatorname{Bound}$ by an algorithm that reads in more and more information about its input, and provides more and more information about its output. We start to copy the Martin-L\"of random $\alpha$ obtained as input to $d_\mlr$ as input to $\Phi$. Whenever we find some $k$ in the input to $\operatorname{Bound}$ while the current prefix of the input to $\Phi$ is $v$, we extend by $w$ as in Lemma \ref{lemma:coaux}, and then continue to write $\alpha$. As the input to $\operatorname{Bound}$ will stabilize, this procedure produces some $\beta \in \mlr$. Moreover, we find that 
if $K \in \operatorname{GreaterNat}(\max(\Phi(\beta)))$, then $K$ is a valid output for $\operatorname{Bound}$.
\end{proof}
\end{thm}

\subsection{Law of the iterated logarithm}
The \emph{law of the iterated logarithm} states that a one-dimensional random walk will eventually remain within a given sublinear (in time) bound around the origin. We consider its effective version:

\begin{defi}
Let $\lil : \mlr \mto \mathbb{N}$ be defined via $N \in \lil(\alpha)$ iff: \[\forall n \geq N \quad  |\sum_{i = 0}^{n-1} (2\alpha(i) - 1)| < \sqrt{2n \log \log n}\]
\end{defi}

It was shown by \name{Vovk} \cite{vovk} that $\lil$ is well-defined, and it is shown in \cite{davie} that $\lil$ is layerwise computable.

\begin{lem}
\label{lemma:auxlil}
Given $N \in \mathbb{N}$ and $u \in \{0,1\}^*$, we can compute some $v \in \{0,1\}^*$ such that $|uv| > N$ and $|\sum_{i=0}^{|uv|-1} (2(uv)(i) - 1)| > \sqrt{2|uv| \log \log |uv|}$.
\begin{proof}
Let $|u| = k$, and assume $v$ is of the form $v = 1^{k+l}$ for some $l \in \mathbb{N}$. Then $|\sum_{i=0}^{|uv|-1} (2(uv)(i) - 1)| \geq l$. Thus choosing $l > N - k$ satisfying $l > \sqrt{2(2k+l) \log \log (2k+l)}$ suffices for our purpose. This in turn can be achieved by $l \geq \max \{20, 2k\}$.
\end{proof}
\end{lem}

\begin{thm}
$\lil \equivW \lay$.
\begin{proof}
The direction $\lil \leqW \lay$ follows from Observation \ref{obs:layer} and the layerwise computability of $\lil$ \cite[Theorem 7]{davie}. For the other direction, we show $d_\mlr \times \operatorname{Bound} \leqW \lil$ instead and employ Theorem \ref{theo:main} and Lemma \ref{lemma:cn}.

The random input to $d_\mlr$ is copied to the input to $\lil$. If a new number $N$ appears in the input to $\operatorname{Bound}$ while the current prefix to the input for $\lil$ is $v$, we extend the input to $\lil$ according to Lemma \ref{lemma:auxlil}. Then we continue to copy over the random input. As the input to $\operatorname{Bound}$ will stabilize eventually, this procedure results in a random input to $\lil$, and by constructing, any output from $\lil$ will be a valid output for $\operatorname{Bound}$.
\end{proof}
\end{thm}

\subsection{Birkhoff's theorem}
The convergence speed in a special case of Birkhoff's theorem was one of the first examples of a layerwise-computable map, already given as such in \cite[Theorem 5.2.4]{hoyrup6} by \name{Hoyrup} and in \cite{galatolo} \name{Rojas} and by \name{Galatolo}, \name{Hoyrup} and \name{Rojas}. Here we shall only consider a toy version -- essentially, the strong law of large numbers in disguise. This toy version already is Weihrauch-complete for layerwise computability, which then of course is inherited by any more general but still layerwise computable versions.

Let $S : \Cantor \to \Cantor$ be the usual shift-operator, and $\pi_1 : \Cantor \to \{0,1\}$ be the projection to the first bit. Let $\operatorname{Birkhoff} : \mlr \times \mathbb{N} \mto \mathbb{N}$ be defined via $N \in \operatorname{Birkhoff}(p,k)$ iff $\forall n \geq N$ we find that: $$\left | \left ( \frac{1}{n+1}\sum_{i= 0}^n \pi_1(S^i(p)) \right ) - \frac{1}{2} \right | < 2^{-k}$$

\begin{lem}
\label{lemma:birkhoffaux}
Given $u \in \{0,1\}^*$ and $k, N \in \mathbb{N}$, $k > 0$, we can compute some $v \in \{0,1\}^*$ such that $|uv| \geq N$ and:
$$\left | \left ( \frac{1}{|uv|}\sum_{i= 0}^{|uv| -1} \pi_1(S^i(uv)) \right ) - \frac{1}{2} \right | > 2^{-k}$$
\begin{proof}
Choosing $v := 0^l$ for sufficiently large $l$ makes the statement true, and we can decide for any value of $l$ whether it is already large enough.
\end{proof}
\end{lem}

\begin{thm}
\label{theo:birkhoff}
$\operatorname{Birkhoff} \equivW \lay$
\begin{proof}
The reduction $\operatorname{Birkhoff} \leqW \lay$ follows from \cite[Theorem 6]{davie} establishing layerwise computability of $\operatorname{Birkhoff}$ and Observation \ref{obs:layer}.

For the reverse direction, we show $d_\mlr \times \operatorname{Bound} \leqW \operatorname{Birkhoff}$ instead, invoking Theorem \ref{theo:main} and Lemma \ref{lemma:cn}. We copy the random sequence provided as input to $d_\mlr$ over to the input for $\operatorname{Birkhoff}$. If some number $N$ is listed in the input to $\operatorname{Bound}$, we extend the current input to $\operatorname{Birkhoff}$ as in Lemma \ref{lemma:birkhoffaux} with $w$ as the current prefix of the input to $\operatorname{Birkhoff}$ and $k = 1$. After that, we proceed to copy the random sequence.

Eventually, the input to $\operatorname{Bound}$ stabilizes, so the input $p$ to $\operatorname{Birkhoff}$ has a random tail and thus is random itself. By construction, if $N \in \operatorname{Birkhoff}(p,1)$, then $N$ is a valid output for $\operatorname{Bound}$.
\end{proof}
\end{thm}

In \cite{hoyrup7,franklin2} it is shown that every element of $\mlr$ satisfies the convergence condition in Birkhoff's ergodic theorem for effectively open respectively effectively closed sets.
In general however, the rate of convergence is not layerwise computable\footnote{A counterexample had already been presented in \cite[Theorem 1]{vyugin}. While the result is formulated in turms of non-effectiveness of convergence in probability, it is easily seen that layerwise computability of the (pointwise) rate of convergence implies effective convergence in probability.}. The lower bound for the Weihrauch degree of finding such a rate of convergence provided in Theorem \ref{theo:birkhoff} of course still applies, but finding upper bounds and a precise classification seems to be an interesting open area.

\subsection{Random harmonic series}
The harmonic series $\sum_{n \in \mathbb{N}} \frac{1}{n}$ might be the most famous example of a diverging series. If, however, the signs of the summands are chosen by independent coin flips, the resulting series will almost-surely converge. Some observations on the resulting distribution can be found in \cite{schmuland}. The effective counterpart was found by \name{Dai}:

\begin{thm}[{\cite[Theorem 2]{dai}, Special case}]
\label{theo:laurent}
The map $p \mapsto \sum_{n \in \mathbb{N}} \frac{(-1)^{p(n)}}{n} : \mlr \to \mathbb{R}$ is well-defined and layerwise computable.
\end{thm}

\begin{thm}
\label{theo:harmonic}
$\left ( p \mapsto \sum_{n \in \mathbb{N}} \frac{(-1)^{p(n)}}{n} : \mlr \to \mathbb{R} \right ) \equivW \lay$
\begin{proof}
The reduction from left to right follows from \cite[Theorem 2]{dai} and Observation \ref{obs:layer}. By Theorem \ref{theo:main} and Lemma \ref{lemma:cn} we can show
\begin{align*}
d_\mlr \times \operatorname{Bound} \leqW \left ( p \mapsto \sum_{n \in \mathbb{N}} \frac{(-1)^{p(n)}}{n} : \mlr \to \mathbb{R} \right )
\end{align*}
for the other direction.

Given some $p \in \mlr$ and an non-decreasing bounded sequence $(a_i)_{i \in \mathbb{N}}$, we will obtain some $q \in \mlr$ by almost copying $p$, but changing finitely many $1$'s to $0$'s such that we can guarantee $\sum_{n \in \mathbb{N}} \frac{(-1)^{q(n)}}{n} \geq \max_{n \in \mathbb{N}} a_n$. For this, we inspect both the sequence $(a_n)_{n \in \mathbb{N}}$ and compute the partial sums $\sum_{n = 0}^N \frac{(-1)^{q(n)}}{n}$ for the output written so far. If for some $N \in \mathbb{N}$ we find that $\sum_{n = 0}^N \frac{(-1)^{q(n)}}{n} < a_N$, then we identify finitely many $j_k > \ldots > j_0 > N$ with $p(j_i) = 1$ and $\sum_{n = 0}^N \frac{(-1)^{q(n)}}{n} + 2\sum_{l = 0}^k \frac{1}{j_l} > a_N + 1$. Such $j_l$ must exist as the harmonic series diverges. We then let $q(m)$ for $N < m \leq j_k$ by $q(m) = 0$ if $m = j_l$ for some $l$ and $q(m) = p(m)$ else.

Let $c = \sum_{n \in \mathbb{N}} \frac{(-1)^{p(n)}}{n}$, let $t$ be such that $\forall T > t \ |\sum_{n = t}^{T} \frac{(-1)^{p(n)}}{n}| < 1$ and $N \geq \max_{n \in \mathbb{N}} a_n$. If the procedure above is triggered $k$ times, then $\sum_{n \in \mathbb{N}} \frac{(-1)^{q(n)}}{n} \geq c + k$ is ensured, as each time the limit is increased by at least $1$. Once $c + k \geq N + 1$, and we have processed $p$ up to at least position $t$, it follows that the procedure cannot be triggered again. Thus, the Hamming distance of $p$ and $q$ is finite, and hence $q \in \mlr$ follows. That the limit satisfies the criterion is immediate.
\end{proof}
\end{thm}

In \cite{dai}, a general result was established regarding when some limit of the form
\begin{align*}
\sum_{n \in \mathbb{N}} (-1)^{p(n)}a_n
\end{align*}
is guaranteed to exist for $p \in \mlr$, and moreover, to be layerwise computable. We point out that the proof of Theorem \ref{theo:harmonic} is not referring to specific properties of the harmonic series beyond its divergence, and hence extends in a straight-forward manner to a more general case.

As a consequence of Theorem \ref{theo:harmonic}, we can find an example for a problem that is layerwise computable, not computable and not Weihrauch complete for layerwise computability. This example was suggested as a promising candidate to the authors by Mathieu Hoyrup and Laurent Bienvenu at CCR 2015.

\begin{cor}
The map $\operatorname{SumApr} : \mlr \times \mathbb{Q} \times \mathbb{N} \mto \{0,1\}$ with $0 \in \operatorname{SumApr}(p,q,k)$ if $\sum_{n \in \mathbb{N}} \frac{(-1)^{p(n)}}{n} < q + 2^{-k}$ and $1 \in \operatorname{SumApr}(p,q,k)$ if $\sum_{n \in \mathbb{N}} \frac{(-1)^{p(n)}}{n} > q$ is
\begin{enumerate}
\item layerwise computable,
\item not computable,
\item not Weihrauch complete for layerwise computability.
\end{enumerate}
\begin{proof}
\begin{enumerate}
\item As a consequence of Theorem \ref{theo:laurent}.
\item If $\operatorname{SumApr}$ were computable, then we could compute the map from Theorem \ref{theo:harmonic} by exhaustive search, contradicting that theorem.
\item It was shown in \cite{paulymaster} that $\C_\mathbb{N}$ is not reducible to any map with finite range even relative to some oracle (i.e.~with continuous instead of computable witness functions $H$,$K$ in Definition \ref{def:weihrauch}). As $d_\mlr \times \C_\mathbb{N}$ is equivalent to $\C_\mathbb{N}$ relative to any ML-random oracle $p \in \mlr$, the claim follows from Theorem \ref{theo:main}.
  \qedhere
\end{enumerate}
\end{proof}
\end{cor}

The generalization from random harmonic series to random Fourier series was explored by \name{Potgieter} \cite{paulpotgieter}, and might provide for further examples of problems that are Weihrauch-complete for layerwise computability.

\section{Hitting time}
\label{sec:hitting}
Natural counterexamples\footnote{The existence of counterexamples, albeit of a more technical nature, is also shown in \cite{shafer2}.} to the converse of Observation \ref{obs:layer} (i.e.~problems that are Weihrauch reducible to $\lay$ but not layerwise computable) are found in hitting time operators. These take an additional input besides the random sequence though, and we need to clarify what layerwise computability means here: A function $f : \mlr \times \mathbf{X} \to \mathbf{Y}$ shall be called \emph{layerwise computable} relative to the universal test $(U_n)_{n \in \mathbb{N}}$, if there is a computable function $F:\subseteq \mlr \times \mathbb{N} \times \mathbf{X} \to \mathbf{Y}$ such that if $p \in \bigcup_{i = 1}^k U_i^C$, then $F(p,k,x) = f(p,x)$ for all $x \in \mathbf{X}$.

Let $T : \Cantor \to \Cantor$ be the usual shift-operator. For some space $\mathcal{P}(\Cantor)$ of subsets of $\Cantor$, we define $\operatorname{HittingTime}_{\mathcal{P}} : \subseteq \mlr \times \mathcal{P}(\Cantor) \to \mathbb{N}$ via $\operatorname{HittingTime}(p,U) = \min \{n \in \mathbb{N} \mid T^n(p) \in U\}$. The two cases we consider is $\mathcal{P} = \mathcal{O}$ and $\mathcal{P} = \mathcal{A}$. It is easy to see that $\operatorname{HittingTime}_{\mathcal{O}}(p,U)$ is defined for all $U \neq \emptyset$. It was shown by \name{Ku\v{c}era} \cite{kucera} that if $p$ is ML random relative to some name of $A \in \mathcal{A}(\Cantor)$ and $A$ has positive measure, then $(p,A) \in \dom(\operatorname{HittingTime}_{\mathcal{A}})$.

\begin{thm}
$\operatorname{HittingTime}_{\mathcal{O}} \equivW d_\mlr \times \lpo^* \leW \lay$, but $\operatorname{HittingTime}_{\mathcal{O}}$ is not layerwise computable.
\begin{proof}
\begin{enumerate}
\item $\operatorname{HittingTime}_{\mathcal{O}} \leqW d_\mlr \times \lpo^*$

Note that $\left (\min :\subseteq \mathcal{O}(\mathbb{N}) \to \mathbb{N} \right ) \equivW \lpo^*$ as shown in \cite{mylatz}. Given $p \in \Cantor$, $U \in \mathcal{O}(\Cantor)$ we can compute $\{n \mid T^n(p) \in U\} \in \mathcal{O}(\mathbb{N})$. The claim follows.

\item $d_\mlr \times \lpo^* \leqW \operatorname{HittingTime}_{\mathcal{O}}$
Again, we use $\left (\min :\subseteq \mathcal{O}(\mathbb{N}) \to \mathbb{N} \right ) \equivW \lpo^*$ and show $d_\mlr \times \min \leqW \operatorname{HittingTime}_{\mathcal{O}}$ instead. Our input is some $p \in \mlr$ and some non-empty $U \in \mathcal{O}(\mathbb{N})$. We inspect $U$ until we find some element $b \in U$ (which provides an upper bound for $\min U$).

We proceed to construct the random sequence $q$ used as the first input to $\operatorname{HittingTime}_{\mathcal{O}}$. For $i \leq b$, let $w_i \in \{0,1\}^{2+2\lceil \log b \rceil}$ be the sequence that starts with $11$ and then intersperses zeros and the digits in a binary code for $i$ of length $\lceil \log b \rceil$, ending with $0$. Then we let $q := w_0w_1\ldots w_bp$.

Next, we construct the open set $V \in \mathcal{O}(\Cantor)$ used as the second input to $\operatorname{HittingTime}_{\mathcal{O}}$. We let $V = \bigcup_{\{i \leq b \mid i \in U\}} w_i\Cantor$. Then we find that for $j \leq (2+2\lceil \log b \rceil)b$ we have $T^j(q) \in V$ iff $j = l(2+2\lceil \log b \rceil)$ and $l \in U$. As we can compute $l$ from $j$ and $b$, the reduction works.

\item $\operatorname{HittingTime}_{\mathcal{O}}$ is not layerwise computable.

If $\operatorname{HittingTime}_{\mathcal{O}}$ were layerwise computable, then for any $p \in \mlr$ the map $U \mapsto \operatorname{HittingTime}_{\mathcal{O}}(p,U)$ would need to be computable. Fix some $p \in \mlr$. We construct some non-empty $U_q \in \mathcal{O}(\Cantor)$ from $p$ and $q \in \Cantor$ by letting $U_q$ accept $r \in \Cantor$ with $r(0) \neq p(0)$ straight-away, and if some $n \in \mathbb{N}$ with $q(n) = 1$ has been found, then all $r \in \Cantor$ are accepted. We thus find that $\operatorname{HittingTime}_{\mathcal{O}}(p,U_q) = 0$ iff $q = 0^\mathbb{N}$ (i.e.~we have exhibited a reduction $\lpo \leqW \left ( U \mapsto \operatorname{HittingTime}_{\mathcal{O}}(p,U) \right )$). This shows that $U \mapsto \operatorname{HittingTime}_{\mathcal{O}}(p,U)$ is not computable. (\footnote{An alternative proof could be obtained by adjusting the argument used to establish the failure of layerwise computability in Theorem \ref{theo:hittingtimesclosed} below.})

\item $d_\mlr \times \lpo^* \leW \lay$

It was shown in \cite{paulymaster} that $\lpo^* \leW \C_\mathbb{N}$ relative to an arbitrary oracle (i.e.~with continuous rather than just computable reduction witnesses) using Hertling's level \cite{hertling}. Essentially, the separation follows from observing that by iteratively removing the points of continuity of $\lpo^*$ from its domain, after $\omega$-many steps the empty set is reached. On the other hand, $\C_\mathbb{N}$ is discontinuous everywhere. This in particular implies that $d_\mlr \times \lpo^* \leW d_\mlr \times \C_\mathbb{N}$, which yields the claim via Theorem \ref{theo:main}.
\qedhere
\end{enumerate}
\end{proof}
\end{thm}

\begin{thm}
\label{theo:hittingtimesclosed}
$\operatorname{HittingTime}_{\mathcal{A}} \equivW \lay$, but $\operatorname{HittingTime}_{\mathcal{A}}$ is not layerwise computable.
\begin{proof}
\begin{enumerate}
\item $\operatorname{HittingTime}_{\mathcal{A}} \leqW \lay$

By Theorem \ref{theo:main}, we can show $\operatorname{HittingTime}_{\mathcal{A}} \leqW d_\mlr \times \C_\mathbb{N}$ instead. By definition, every instance $(p,A)$ to $\operatorname{HittingTime}_{\mathcal{A}}$ computes a Martin-L\"of random $p$. It only remains to prove that $\operatorname{HittingTime}_{\mathcal{A}} \leqW \C_\mathbb{N}$. Let $B \in \mathcal{O}(\mathbb{N})$ be defined as $B = \{N \in \mathbb{N} \mid \forall n < N \ T^np \notin A\}$. We can compute $B$ from $p$ and $A$ (using standard properties of the constructions of $\mathcal{A}(-)$ and $\mathcal{O}(-)$), and $B$ is guaranteed to be non-empty and finite. We then apply $\max : \subseteq \mathcal{O}(\mathbb{N} \to \mathbb{N}$ (which is equivalent to $\C_\mathbb{N}$ by Lemma \ref{lemma:cn}) and to obtain the correct answer to $\operatorname{HittingTime}_{\mathcal{A}}(p,A)$.

\item $\lay \leqW \operatorname{HittingTime}_{\mathcal{A}}$

Instead, we show $d_\mlr \times \operatorname{Bound} \leq_W \operatorname{HittingTime}_{\mathcal{A}}$ and use Lemma \ref{lemma:cn}. Starting with $p \in \mlr$ and some $q \in \Baire$ s.t.~$\exists N \in \mathbb{N} \ \{0,\ldots,N\} = \{q(i) \mid i \in \mathbb{N}\}$, we wish to compute some $A \in \mathcal{A}(\Cantor)$ s.t.~$\forall i \in \mathbb{N} \ T^{q(i)}p \notin A$ but $(p,A) \in \dom(\operatorname{HittingTime}_{\mathcal{A}})$.

Given $p \in \Cantor$ and $i < j \in \mathbb{N}$, let $p_{[i\leq j]} \in \{0,1\}^{j-i}$ denote the subword of $p$ from position $i$ to position $j$. Now, simply set $A = \left (\bigcup_{i \in \mathbb{N}} p_{[q(i)\leq 2q(i) + 1]}\Cantor \right )^C$. This is a closed set computable from $p$ and $q$, and by construction satisfies our first criterion. For the second criterion, we note that $\sum_{i \in I} 2^{-i-1} < 1$ for any finite set $I \subset \mathbb{N}$. Thus, there is some $w \in \{0,1\}^*$ with $w\Cantor \subseteq A$, and as $p \in \mlr$, we know that $w$ appears somewhere as a subword in $p$.

\item $\operatorname{HittingTime}_{\mathcal{A}}$ is not layerwise computable.

For $k \in \mathbb{N}$, let $A_k \in \mathcal{A}(\Cantor)$ be the set of all sequences whose prefix of length $k$ is not $1$-compressible. We note that $A_k$ is computable uniformly in $k$, and further note that $(p,A_k) \in \dom(\operatorname{HittingTime}_{\mathcal{A}})$ for any $p \in \mlr$, as any ML random $p$ contains every possible subword of length $k$, including some incompressible ones.

Now assume that $\operatorname{HittingTime}_{\mathcal{A}}$ were layerwise computable, witnessed by some computable $F : \subseteq \mlr \times \mathbb{N} \times \mathcal{A}(\Cantor) \to \mathbb{N}$. We consider the computable (uniformly in $k$) maps $F_k : \subseteq \mlr \to \mathbb{N}$ defined by $F_k(p) = F(p,2k,A_k)$. Using $F_k$ as a subroutine, we will search for some $w \in \{0,1\}^k$ such that for some set $B$ of measure at least $\frac{1}{2}$ we can confirm that $\forall p \in B \ F_k(wp) = 0$. Since there must be some $p \in B$ with $wp \in \bigcup_{i = 1}^{2k} U_i^C$, this implies that $wp \in A_k$, i.e.~that $w$ is not $1$-compressible.

Putting together the pieces, we would have an algorithm that reads some $k \in \mathbb{N}$ and outputs a $1$-incompressible word of length $k$. This is clearly a contradiction, hence $\operatorname{HittingTime}_{\mathcal{A}}$ cannot be layerwise computable.
\qedhere
\end{enumerate}
\end{proof}
\end{thm}

Complementing the results above, the map $\operatorname{HittingTime}_{\mathcal{A} \wedge \mathcal{O}}$ where the set-input is demanded to be clopen (by providing both a name for it as a closed set, and a name for it as an open set) is easily seen to be computable.

\section*{Acknowledgements}
The work has benefited from the Marie Curie International Research Staff Exchange Scheme \emph{Computable Analysis}, PIRSES-GA-2011- 294962. The second author was supported by the National Research Foundation (NRF) of South Africa.

\bibliographystyle{plainurl}
\bibliography{layerwise}

\end{document}